\documentstyle[preprint,aps,epsf,floats]{revtex}

\begin{document}
\setlength\baselineskip{15pt}

%\preprint{\tighten \vbox{\hbox{WSUTC-ADJ-01-0001}
%        \hbox{hep-th/0000000} 
%}}

\title{Pedagogical Reflections on Color Confinement in Chromostatics \\  }
\author{Richard S. Wittman\footnote{rwittman@tricity.wsu.edu}}
\address{\tighten {\it Washington State University Tri-Cities\\
2710 University Drive\\
Richland, WA 99352-1671}}
\maketitle
\vspace*{-.1in}
\begin{center} 
%(\today)
(January 2001)
\end{center}
\vspace*{-.4in}
{\tighten
\begin{abstract}
Abelian and nonabelian gauge invariant
states are directly compared to revisit how 
the unconfined abelian theory is expressed.
It is argued that the Yang-Mills equations have no obvious 
physical content apart from their relation to underlying 
physical states.  
The main observation is that the physical states of 
electrostatics can be regarded as point charges connected 
by a uniform superposition of all possible Faraday lines.  
These states are gauge invariant only in the abelian
case.\\ \\
%PACS number(s): 12.38.Aw
\end{abstract}
}%end tighten

%\section{ INTRODUCTION}
After the initial wonder and awe that students experience on
learning  that the nuclear force can be fundamentally 
described with a nonabelian gauge theory, there is a natural 
desire to solve the equations---that is the Yang-Mills equations.  
There is a naive expectation
that we can calculate a chromostatic potential energy for an 
arrangement of quark color charge in the same way as for electric 
charge in the Maxwell equations.  Ideally, we expect to find a 
confining potential that would explain why unbound quarks
possessing the color charge are
not observed.  Additionally, the conditions under which 
deconfinement could occur should somehow emerge from the 
equations.   

Unfortunately, it seems that the simplest solutions have no 
relevance to physical hadrons.  General solutions to the nonlinear 
field equations cannot be obtained from linear combinations of 
known solutions.  Going to the quantized theory this implies that 
free quanta of the gauge fields (gluons) do not exist in the same 
sense as photons.  In trying to understand this 
situation, a number of plausible explanations of color confinement
have been studied.  For examples, see \cite{GtH,Magpantay,RMW,Hansson}.  
It has even been suggested that the confinement problem itself 
can be eliminated by recognizing that it is an artifact of false
expectations of nonabelian gauge theories\cite{Hansson}.  
At least at low energies, Feynman diagrams with gluon 
exchange are conceptually misleading and physically wrong when
taken individually.

Therefore, we first want to show how the naive expectation
of calculating a classical potential goes wrong.  
Then we try to understand color confinement as well 
as what is meant by deconfinement by directly comparing 
the nonabelian and abelian cases in the simplest 
situation.  A way to make the comparison direct, is to consider 
the abelian case as an $SO$(2) projection of $SO$(3).  The gauge 
transformations in the abelian theory are then viewed as local 
rotations about the third axis---quantities 
carrying the third
free index are invariant with respect to such rotations.

Consider the Yang-Mills Gauss' law for static color electric fields
${\bf E}^b({\bf x})$ and vector potentials ${\bf A}^b({\bf x})$ 
($b=1$,2 or 3):
\begin{eqnarray}
&\displaystyle{ 
\nabla \cdot {\bf E}^a + g
\epsilon_{abc}{\bf A}^b \cdot {\bf E}^c =
\rho^a
}\label{gauslaw}\end{eqnarray}
where $g$ is the coupling constant and the 
antisymmetric $\epsilon_{abc}$ gives the $SU$(2) structure 
constants (summation over the repeated indices is implied).
The color electric energy associated with charge distribution $\rho^a({\bf x})$
is
\begin{eqnarray}
&\displaystyle{ 
V =
\frac{1}{2}\int\!\!d^3x\,
{\bf E}^a \cdot {\bf E}^a \,\, .
}\label{staticenergy}\end{eqnarray}
Also, Eq.~(\ref{gauslaw}) is gauge covariant and $V$ is gauge invariant
with respect to
the local $SO$(3) transformations:
\begin{eqnarray}
&\displaystyle{ 
\rho'^a =  R_{ab} \rho^b,\,\,\,\,\,{\bf E'}^a = R_{ab} {\bf E}^b
}\label{gaugetransformation1}\end{eqnarray}
and
\begin{eqnarray}
&\displaystyle{ 
{\bf A'}^a = R_{ab} {\bf A}^b - \frac{1}{2g}\epsilon_{abc} R_{bd} \nabla R_{cd}
}\label{gaugetransformation2}\end{eqnarray}
where $R_{ab}({\bf x})R_{ac}({\bf x}) = \delta_{bc}$.

A standard approach of electrostatics would be to solve the differential 
equation (Eq.~(\ref{gauslaw})) and evaluate the integral of 
Eq.~(\ref{staticenergy}). Of course this approach gives the 
Coulomb potential for point charges $\rho^3({\bf x}) = 
q_1\delta^3({\bf x}-{\bf x}_1) + 
q_2\delta^3({\bf x}-{\bf x}_2)$ in the abelian case where
all the indices in Eqs.~(\ref{gauslaw}) and~(\ref{staticenergy}) 
are replaced with a 3.  For the nonabelian case we are immediately
faced with a puzzle---physically, what is $\rho^a$?  Even for 
the simplest situation in the classical theory there is
already a puzzle.  For instance, trying to define two point
charges according to $\rho^a({\bf x}) = 
q^a_1\delta^3({\bf x}-{\bf x}_1) + 
q^a_2\delta^3({\bf x}-{\bf x}_2)$ is not physically meaningful
because $q^a_1$ and $q^a_2$ can be arbitrarily and independently
rotated with a gauge transformation.  One could argue that
specifying $q^a_1$, $q^a_2$ and ${\bf A}^b({\bf x})$
together would be mathematically sufficient to determine $V$, 
but the physical significance of $V$ would still be unclear without 
some physical gauge invariant entity (such as electric charge). 
It is true that $\Delta V$ is the work required to change between 
different spatial configurations of $\rho^a$ and ${\bf A}^b$,
but spatial configurations of what?  It's as if we are given
a geographer's data on coastline locations that are hopelessly 
entangled with a redefinition of latitude and longitude at each
point on the map.  The observable changes (in climate, vegetation, {\it etc}.) 
associated with each data set might be known, but the existence
of actual coastlines is hidden.  Calculating the
energy associated with different data sets i.e.~$\rho^a$ and 
${\bf A}^b$ does not solve the puzzle of what physically
exists within data.  In fact, it is probably misleading
because if we set ${\bf A}^b = {\bf 0}$ and $\rho^a({\bf x}) = 
q^a_1\delta^3({\bf x}-{\bf x}_1) + q^a_2\delta^3({\bf x}-{\bf x}_2)$
we obtain the Coulomb potential as in electrostatics--yet this time
with nothing that can legitimately be identified as physical sources.
In a similar sense, recall that the local gravitational energy 
density is not a meaningful observable because it always can be
transformed to zero with local geodesic coordinates\cite{MTW}.
Therefore, instead of approaching chromostatics in the same spirit
as electrostatics, let's turn the problem around by respecting the
nonabelian gauge invariance of physical states and then show 
how electrostatics is recovered for the abelian theory.

To begin, let's take the point of view that the nonabelian 
charge density of Eq.~(\ref{gauslaw}) only makes sense in
terms of an operator acting on some physical state.  The 
simplest choice for our discussion is to work with
charges in the adjoint representation of $SU$(2)
\begin{eqnarray}
&\displaystyle{ 
\rho^a({\bf x}) = g 
\,a_b^\dagger({\bf x}) T^a_{bc} a_c({\bf x})
}\label{charge}\end{eqnarray}
where $T^a_{bc}=i\epsilon_{bac}$ and 
$a_b^\dagger$ and $a_c$ are the creation and annihilation operators 
for point charges with respect to a gauge invariant state $|0\rangle$ 
such that $a_c|0\rangle = 0$.  Also, the operators satisfy the 
commutation relations
\begin{eqnarray}
&\displaystyle{
[a_a({\bf x}),a_b^\dagger({\bf x}')]= \delta_{ab}\delta^3({\bf x}-{\bf x}')\,\,.
}\label{aadagcom}\end{eqnarray}
Since the creation and annihilation operators transform the same
way as the quantities in Eq.~(\ref{gaugetransformation1}), 
their color indices must be contracted locally to form a gauge
invariant operator.  
Geometrically, the simplest gauge invariant
state contains two point charges and is created by the operator
\begin{eqnarray}
&\displaystyle{
\Pi [\gamma] =
a_a^\dagger({\bf x}(1))
U_{ab} [\gamma] a_b^\dagger({\bf x}(0))\,\,.
}\label{twocharges}\end{eqnarray}
Such a state represents spatially separated point charges at ${\bf x}(0)$ 
and ${\bf x}(1)$ connected with the Wilson line along some curve 
$\gamma: {\bf x} = {\bf x}_\gamma(s)$.  The parameterization used for
any curve $\gamma$ can arbitrarily be taken on the interval $0\le s \le 1$.
The path ordered exponential $U_{ab} [\gamma]$ is defined according to
\begin{eqnarray}
&\displaystyle{
\left(
\delta_{ab} \frac{d}{ds} -ig
T^c_{ab}{\bf A}^c \left({\bf x}_{\gamma}(s)\right) \cdot 
\frac{d{\bf x}_{\gamma}(s)}{ds}
\right) U_{bd} \left[\gamma_0^s\right] = 0
}\label{wlinedef}\end{eqnarray}
where
\begin{eqnarray}
&\displaystyle{
U_{ab} [\gamma_0^0] = \delta_{ab} \,\,\mbox{ \rm and}\,\,\,
U_{ab} [\gamma] \equiv U_{ab} [\gamma_0^1]\,\,\,.
}\label{wlinedef0}\end{eqnarray}
Notice that a similar $\Pi [\gamma]$ creates a gauge
invariant two charge state in the abelian case.  
The corresponding abelian forms of Eqs.~(\ref{charge}-\ref{wlinedef0}) 
are obtained by restricting all the lower color indices to 1 or 2, 
or equivalently by inserting the projection $(T^3T^3)_{ab}$ between
each matrix product.  Therefore, even for the abelian case, the physical
charges of a gauge invariant state require the nonlocal presence of
field lines.  As thoroughly explained in a series of 
papers by Lavelle and McMullan (for example see \cite{LM}), this 
is the point of view of Dirac\cite{Dirac}. Dirac showed that
the operator that creates a physical electron state, must create
an electron together with its Coulomb field. 

In order to satisfy Gauss' law of Eq.~(\ref{gauslaw}) with an 
operator valued source we require that
\begin{eqnarray}
&\displaystyle{ 
\left[
\nabla \cdot {\bf E}^a +
g\epsilon_{abc}{\bf A}^b \cdot {\bf E}^c -
g a_b^\dagger({\bf x}) T^a_{bc} a_c({\bf x})
,\Pi[\gamma]\right] = 0
}\label{opgauslaw1}\end{eqnarray}
which implies the following commutation relation for the 
path ordered exponential
\begin{eqnarray}
&\displaystyle{
\left[
\nabla \cdot {\bf E}^a({\bf x}) +
g\,\epsilon_{abc}{\bf A}^b({\bf x}) \cdot {\bf E}^c({\bf x})
,\,U_{mn} [\gamma]
\right]= \phantom{OOOOOOOOOOOOOO}
}\nonumber \\&\displaystyle{\phantom{OOOOOOOOO}
  i g \delta^3\left({\bf x}- {\bf x}(0)\right)
  U_{mb} [\gamma] \epsilon_{abn}
- i g \delta^3\left({\bf x}- {\bf x}(1)\right)
  \epsilon_{amc} U_{cn} [\gamma]  \,\,.
}\label{opgauslaw2}\end{eqnarray}
It can be shown that Eq.~(\ref{opgauslaw2}) follows from the definition
of Eq.~(\ref{wlinedef}) if we assume that
\begin{eqnarray}
&\displaystyle{
\left[
{\bf E}^a({\bf x}),U_{mn} [\gamma]
\right]= i g
 \int_0^1\!\!ds\,
\frac{d{\bf x}_{\gamma}(s)}{ds}\,
\delta^3\left({\bf x}- {\bf x}_{\gamma}(s)\right)
  U_{mb} [\gamma_s^1] iT^a_{bc}
 U_{cn} \left[\gamma_0^s\right]\,\,.
}\label{Ecommutator}\end{eqnarray}
It is interesting to notice that Eq.~(\ref{Ecommutator}) would
follow directly from Eq.~(\ref{wlinedef}) by assuming the
canonical relations $[E^a_j({\bf x}),A^b_k({\bf x}')]=i\delta_{jk}
\delta_{ab}\delta^3({\bf x}-{\bf x}')$.  This is consistent with
the fact that Eq.~(\ref{opgauslaw1}) is a statement of the gauge 
invariance of $\Pi[\gamma]$. The fact that the chromostatic 
fields ${\bf E}^a$ are not directly observable because
they are not gauge invariant suggests that Eq.~(\ref{Ecommutator})
is as far as we can go in terms of {\it solving} Eq.~(\ref{gauslaw}).
We get a gauge invariant expression by contracting the color
indices in a double commutator using Eq.~(\ref{Ecommutator}) 
twice to get 
\begin{eqnarray}
&\displaystyle{
\left[
E^a_j({\bf x}), 
\left[
E^a_j({\bf x}),\Pi [\gamma]
\right]
\right]= 2 
\left|
g\bbox{\Delta}\left({\bf x},\gamma \right)
\right|^2
 \Pi [\gamma]
}\label{Ecommutator2}\end{eqnarray}
where the factor of two on the right is the $SO$(3) representation dependent
factor $L(L+1)$ with $L=1$.  For $SO$(2) the factor itself is simply one.
Also, the vector quantity on the right can be regarded as a single
Faraday line\cite{MF} of the field:
\begin{eqnarray}
&\displaystyle{
\bbox{\Delta}\left({\bf x},\gamma \right) \equiv
\int_0^1\!\!ds\,
\frac{d{\bf x}_\gamma(s)}{ds}\,
\delta^3\left({\bf x}- {\bf x}_\gamma(s)\right)\,\, .
}\label{fluxtube}\end{eqnarray}
Assuming a gauge invariant state $|0\rangle$ exists such that 
$E^a_j|0\rangle=0$,
Eq.~(\ref{Ecommutator2}) can be used to show that
$\Pi[\gamma]|0\rangle$ is an eigenstate of 
${\bf E}^a \cdot {\bf E}^a$ and is therefore an
eigenstate of $V$
\begin{eqnarray}
&\displaystyle{
V\,\,\Pi[\gamma]|0\rangle =
\frac{1}{2}\int\!\!d^3x\,
2\left|
g\bbox{\Delta}\left({\bf x},\gamma \right)
\right|^2
\Pi[\gamma]|0\rangle
}\label{eeeigenvalue}\end{eqnarray}
where the integral is proportional to the 
length of the curve $\gamma$.  Although it is
divergent, it can be rendered finite by 
absorbing a divergent factor into $g$.
We have only verified that the state
representing a single flux line gives a linear confining potential.
No surprise---this is the most basic result of strong coupling 
on the lattice\cite{KS} and of string models of hadrons\cite{YN,EW}.  
What might be a surprise, though, is that the abelian result also 
gives a linear confining potential that is identical 
to Eq.~(\ref{eeeigenvalue}), but without the factor of two 
in the integrand.  The fact that confinement occurs for both 
cases is conventionally understood as the strong coupling limit 
result of neglecting the magnetic energy\cite{KS,EW}. 
This is true in the nonabelian case but, notice that the key 
difference in the abelian case is that we can go back to  
Eq.~(\ref{Ecommutator}) to use the fact that $T^3$ now commutes 
with $U$.  Two linear combinations of $U$ and $T^3U$ allow us to
obtain ${\bf E}^3$ as an observable directly
\begin{eqnarray}
&\displaystyle{
\left[
{\bf E}^3({\bf x}),\left(U [\gamma]
\mp T^3 U [\gamma]\right)_{mn}
\right]=  
\pm g\bbox{\Delta}\left({\bf x},\gamma \right)
\left(  U [\gamma] \mp T^3 U [\gamma]
\right)_{mn}
}\label{abeleancommutator}\end{eqnarray}
where the linear combination with the $-$($+$) sign corresponds
to a positive (negative) charge at ${\bf x}(0)$ and a 
negative (positive) charge at ${\bf x}(1)$.  This is dramatically
different from working with Eq.~(\ref{Ecommutator2}) where
the coherence of the field lines has been lost.
%This state might provide the more natural regulator for
%the divergent integral in Eq.~(\ref{eeeigenvalue}).
Operators involving products of $\Pi[\gamma]$
corresponding to more charges give an incoherent sum of 
squares of Faraday lines.   The abelian theory is
completely different at this point; because the electric field
is an observable, a state corresponding to a coherent
superposition of Faraday lines can be constructed with
\begin{eqnarray}
&\displaystyle{
P_\pm [f] =
a^\dagger_m({\bf x}(1)) \left\{ \prod_{\gamma}
\left(  U [\gamma] \mp T^3 U [\gamma]
\right)^{f[\gamma]} \right\}_{mn} a^\dagger_n({\bf x}(0))
}\label{prodstate}\end{eqnarray}
where a formal product over curves between ${\bf x}(0)$
and  ${\bf x}(1)$ is represented.
Notice that even though Eq.~(\ref{prodstate}) involves a
nonlocal contraction of indices within the product,
it is still gauge invariant if
\begin{eqnarray}
&\displaystyle{
\sum_{\gamma} f[\gamma] = 1\,\,\, .
}\label{normsum}\end{eqnarray}
This would be more obvious for the analogous
$U$(1) expression in terms of complex numbers instead
of commuting matrices.  Consistent with Eq.~(\ref{prodstate}),
the commutator
\begin{eqnarray}
&\displaystyle{
\left[
{\bf E}^3({\bf x}),P_\pm [f] \right] = 
\pm g \left\{ \sum_\gamma
\bbox{\Delta}\left({\bf x},\gamma \right)
f[\gamma] \right\}P_\pm[f]
}\label{acommutator}\end{eqnarray}
involves a formal average over Faraday lines between ${\bf x}(0)$
and  ${\bf x}(1)$.  {\it It is this coherent average that allows
deconfinement in the abelian case}.  We could speculate here that
such a state could also arise in the nonabelian theory
where some mechanism provides the appropriate abelian 
projections.  
Recall that for us, inserting 
$(T^3T^3)_{ab}$ by hand in the nonabelian
expressions gives the abelian expressions.

The average of Eq.~(\ref{acommutator}) can be made explicit
by evaluating a ``sum over curves'' path integral
\begin{eqnarray}
&\displaystyle{
\left\langle\!\left\langle 
\bbox{\Delta}\left({\bf k},\gamma \right)
\right\rangle\!\right\rangle_\Lambda=
{\cal N}\int_\gamma {\cal D}\left[{\bf x}_\gamma(s)\right]
\exp\left[ -\frac{\Lambda}{2}\int_0^1\!\!ds
\left( \frac{d{\bf x}_\gamma(s)}{ds} \right)^2
 \right]
\bbox{\Delta}\left({\bf k},\gamma \right)
}\label{defcurveavg}\end{eqnarray}
where we work with the Fourier transformed Faraday line  
\begin{eqnarray}
&\displaystyle{
\bbox{\Delta}\left({\bf k},\gamma \right) =
\int_0^1\!\!ds\,
\frac{d{\bf x}_\gamma(s)}{ds}
\exp\left(
-i{\bf k} \cdot {\bf x}_\gamma(s)
\right)
}\label{kcurve}\end{eqnarray}
and of course ${\cal N}$ is defined by the normalization condition
\begin{eqnarray}
&\displaystyle{
{\cal N}\int_\gamma {\cal D}\left[{\bf x}_\gamma(s)\right]
\exp\left[ -\frac{\Lambda}{2}\int_0^1\!\!ds
\left( \frac{d{\bf x}_\gamma(s)}{ds} \right)^2
 \right] = 1 \,\, .
}\label{normalization}\end{eqnarray}
The average in Eq.~(\ref{defcurveavg}) corresponds to a special 
case of the Eq.~(\ref{acommutator})
average where a single parameter $\Lambda$ regulates the scale for
which the curves extend throughout space.  As $\Lambda$ goes to zero
all curves throughout space are equally weighted; as $\Lambda$ 
goes to infinity only the line directly between the charges
contributes. It is tempting to interpret $\Lambda$ in terms of
a string tension, but since the line integral in the exponent 
in Eq.~(\ref{defcurveavg}) is not reparameterization
invariant we will resist the temptation.  The practical
reason for the choice of the exponent
is that it is translationally invariant and can be explicitly
computed from the generating functional
\begin{eqnarray}
&\displaystyle{ {\cal Z}\left[
{\bf x}(0),{\bf x}(1),{\bf J}(s),\Lambda \right] = 
\int_\gamma {\cal D}\left[{\bf x}_\gamma(s)\right]
\exp\left\{ -\int_0^1\!\!ds\left[\frac{\Lambda}{2}
\left( \frac{d{\bf x}_\gamma(s)}{ds} \right)^2
-{\bf J}(s) \cdot {\bf x}_\gamma(s)
 \right] \right\} \,\,.
}\label{curvesum}\end{eqnarray}
Evaluating the gaussian integrals of Eq.~(\ref{curvesum}) gives
\begin{eqnarray}
&\displaystyle{ {\cal Z}\left[
{\bf x}(0),{\bf x}(1),{\bf J}(s),\Lambda \right] = 
\exp\left\{\frac{1}{2\Lambda}\int_0^1\!\!ds\int_0^1\!\!dt\,
{\bf J}(s) \cdot {\bf J}(t) \left[
t\,\theta(s-t) \phantom{OOOOOOOO}
\right.\right.}\nonumber \\&\displaystyle{ \phantom{OOOO}
 \left.\left.
+ s\,\theta(t-s) - st \right]
+ \int_0^1\!\!ds\, {\bf J}(s) \cdot \left[
(1-s){\bf x}(0)
+s\,{\bf x}(1)\right]
-\frac{\Lambda}{2}
\left({\bf x}(1) - {\bf x}(0)\right)^2
\right\} \,\,.
}\label{curvesum2}\end{eqnarray}
It is then straight forward using Eq.~(\ref{curvesum2})
to compute the average in Eq.~(\ref{defcurveavg}) to be
\begin{eqnarray}
&\displaystyle{
\left\langle\!\left\langle 
\bbox{\Delta}\left({\bf k},\gamma \right)
\right\rangle\!\right\rangle_\Lambda=
\int_0^1\!\!ds\,\left[\frac{i{\bf k}}{\Lambda}(s-\frac{1}{2})
+ {\bf x}(1) - {\bf x}(0)\right]\phantom{OOOOOOOOOOOOOOOOO}
}\nonumber \\&\displaystyle{\phantom{OOOOOOOO}
\times\exp\left\{ 
-i{\bf k} \cdot \left[
(1-s){\bf x}(0)
+s\,{\bf x}(1)
\right] - \frac{s(1-s)}{2\Lambda}{\bf k}^2
\right\}\,\,.
}\label{curveavg}\end{eqnarray}
Also, it is reassuring to notice that
\begin{eqnarray}
&\displaystyle{ i{\bf k} \cdot 
\left\langle\!\left\langle 
\bbox{\Delta}\left({\bf k},\gamma \right)
\right\rangle\!\right\rangle_\Lambda=
\exp\left( -i{\bf k} \cdot {\bf x}(0)
\right)
-\exp\left( -i{\bf k} \cdot {\bf x}(1)
\right) \,\, ,
}\label{cconserve}\end{eqnarray}
meaning that Gauss' law for point charges
is satisfied by Eq.~(\ref{curveavg}) for any $\Lambda$.
In the limit as $\Lambda\rightarrow 0$, the integral of 
Eq.~(\ref{curveavg}) is dominated by contributions 
near the limits of integration and becomes
the Fourier transform of the familiar Coulomb field
\begin{eqnarray}
&\displaystyle{ 
\lim_{\Lambda\rightarrow 0}
\left\langle\!\left\langle 
\bbox{\Delta}\left({\bf k},\gamma \right)
\right\rangle\!\right\rangle_\Lambda=
\frac{i{\bf k}}{k^2}\left[
\exp\left( -i{\bf k} \cdot {\bf x}(1)
\right)
-\exp\left( -i{\bf k} \cdot {\bf x}(0)
\right)
\right]\,\,.
}\label{curvelimit}\end{eqnarray}
Therefore, we find that the physical state $P_\pm [f] |0\rangle$
of two charge electrostatics that gives 
\begin{eqnarray}
&\displaystyle{
{\bf E}^3({\bf x})\,\,P_\pm [f] |0\rangle = 
\frac{\pm g}{4\pi} \left\{
\frac{{\bf x}-{\bf x}(0)}{\left|{\bf x}-{\bf x}(0)\right|^3}
-
\frac{{\bf x}-{\bf x}(1)}{\left|{\bf x}-{\bf x}(1)\right|^3}
\right\}
P_\pm [f] |0\rangle
}\label{allcurves}\end{eqnarray}
is the one where the curve weight factors $f[\gamma]$ 
are all equal.  It is remarkable that the physical electric
field can be viewed as a uniform collection of all 
possible Faraday lines.  It is clear that this state is
not available to the symmetry unbroken nonabelian theory
because it is built from a product of path ordered exponentials
that is gauge invariant only for the abelian case.  Furthermore,
we expect that deconfinement requires some mechanism to provide 
appropriate abelian projection operators that allow the nonabelian
theory to make use of such a state.  For instance, consider the 
operator $(\phi^aT^a\phi^bT^b)_{mn}$ containing a Higgs field $\phi^a$
that has uniformly fallen in the arbitrary color direction
along the third axis.  For the symmetry broken ground state,
the operator becomes $|\phi_0|^2(T^3T^3)_{mn}$ which is precisely
the projection operator we have used to convert  
nonabelian expressions to abelian ones.

%\section{DISCUSSION}

In summary, a way has been presented in which color confinement can
be understood as a restriction imposed by nonabelian gauge
invariance on the use of states representing a coherent
superposition of color electric flux lines.
Deconfinement is realized in the abelian
theory by a state that represents charges joined by a coherent
average of all possible flux lines.  Therefore, a practical 
view of confinement in the nonabelian case 
is that this state is not allowed.  It is conceivable
that operators resulting from symmetry breaking could 
act as an abelian projection that makes such a state
allowable and results in the deconfinement of color
charge.  Possibly, the {\it deconfinement} of weak isospin
in electroweak theory could be understood in this
way. 

For the symmetric nonabelian case of $SU$(2) charge
without the color magnetic interaction, 
the only consistent static states are 
one-dimensional idealized flux lines that join the 
nonabelian charges\cite{KS,EW}.
It can be shown that such states are eigenstates of $V$
as long as flux lines do not intersect.  Intersections
result in off diagonal interactions that allow flux
lines to re-associate with a new pairing of charges
as in string-flip models.  It may be possible 
that string-flip models can be understood from the 
static implications of gauge invariance.  
To make this connection it would be 
necessary to consider the more realistic case of 
spin-$\frac{1}{2}$ quarks in the fundamental 
representation of $SU$(3).  Ideally, the color 
magnetic energy and dynamical considerations 
should be included in some approximation.

\end{document}